\documentclass[traditabstract]{aa}
\usepackage{graphicx}
\usepackage{natbib}
\bibpunct{(}{)}{;}{a}{}{,}

\newcommand{\mdot}{\dot{M}}
\newcommand{\saxj}{SAX J1808.4--3658}
\newcommand{\xmm}{\it XMM-Newton}
\def\ltsima{$\; \buildrel < \over \sim \;$}
\def\simlt{\lower.5ex\hbox{\ltsima}}
\def\gtsima{$\; \buildrel > \over \sim \;$}
\def\simgt{\lower.5ex\hbox{\gtsima}}

\begin{document}

\title{XMM-Newton detects a relativistically broadened iron line in the spectrum of the ms X-ray pulsar  {\saxj}}

\author{A.Papitto $^{1,2}$\thanks{mail to papitto@oa-roma.inaf.it} \and T.Di Salvo $^{3}$\and A.D'A\`i $^{3}$\and R.Iaria $^{3}$\and L.Burderi $^{4}$\and A.Riggio $^{4}$\and M.T.Menna $^{2}$\and N.R.Robba $^{3}$}

\institute{Dipartimento di Fisica, Universit\`a degli Studi di Roma 'Tor Vergata', 
via della Ricerca Scientifica 1,00133 Roma,Italy \and INAF Osservatorio Astronomico di Roma, via Frascati 33, 
Monteporzio Catone, 00040, Italy \and Dipartimento di Scienze Fisiche ed Astronomiche, 
Universit\`a di Palermo, via Archirafi 36, Palermo, 90123, Italy \and Dipertimento di Fisica, Universit\`a degli Studi di Cagliari, SP Monserrato-Sestu, KM 0.7, Monserrato, 09042 Italy}

\abstract
{
We report on a 63-ks long {\xmm} observation of the accreting millisecond pulsar
SAX J1808.4-3658 during the latest X-ray outburst which started on September
21st 2008. The pn spectrum shows a highly significant emission line in the
energy band where the iron K-$\alpha$ line is expected, and which we identify as
emission from neutral (or mildly ionized) iron. The line profile appears to be
quite broad (more than 1 keV FWHM) and asymmetric; the most probable explanation
for this profile is Doppler and relativistic broadening from the inner accretion
disc. From a fit with a diskline profile we find an inner radius of the disc of
$8.7^{+3.7}_{-2.7}$ R$_g$, corresponding to $18.0^{+7.6}_{-5.6}$ km for a 1.4
M$_{\odot}$ neutron star. The disc therefore appears truncated inside the
corotation radius (31 km for SAX J1808.4-3658) in agreement with the fact that
the source was still showing pulsations during the XMM-Newton observation.
}

\titlerunning{Relativistic Iron Line from {\saxj}}
\authorrunning{A.Papitto et al.}
\maketitle

\section{Introduction}

Accreting millisecond pulsars (AMSP) are generally interpreted as the
evolutionary link between the low mass X-ray binaries (LMXB) and rotation
powered millisecond pulsars. According to the recycling scenario, the latter are
in fact formed after a phase of mass accretion from a low mass companion star,
which ultimately spins the neutron star (NS) up to ms periods \citep[see e.g.][]{BvdH91}.
{\saxj} was the first AMSP discovered \citep{WvdK98}. In ten
years the population of AMSP has grown to ten sources, all residing in close and
transient binaries, but {\saxj} can still be considered the cornerstone of its
class as it has repeatedly gone into outburst almost every two years since its
first detection, making it the most observationally rich source of its class.
Its timing and orbital properties have been extensively studied and are still
debated to some extent (\citet{B06}, \citet{dS08}, \citet{H08},
H08 hereinafter). Its broadband X-ray spectral emission during the outburst
phase was studied using {\it RXTE} observations of the 1998 outburst (\citet{G98},
\citet{HS98}, \citet{GDB02}, GDB hereinafter).

GDB decomposed the 3.0­--150.0 keV X-ray spectrum into a soft component,
originating either from the disc or the NS surface, and a dominating hard
component extending to energies up to 100 keV ($kT_e>35$ keV), interpreted as
Comptonization of soft photons, probably in the accretion column above the
magnetic caps of the NS. A disc reflection component was also observed producing
the typical hump at 30 keV and a K$\alpha$ iron line at $\sim$ 6.5 keV.

Broad iron emission lines, in the 6.4­--6.97 keV energy range, have been
observed from many  NS LMXB (\citet{dS05}, \citet{BS07}, \citet{C08}, \citet{pandel08}, \citet{DA08},
and references therein). Their  broadening  and in
particular the asymmetry  of the line is generally  interpreted as the
effect of the relativistic Keplerian  motion of the reflecting plasma
in an optically thick, and geometrically  thin, accretion disc, in the deep
gravitational well of the nearby compact object \citep{F89}.

The  determination  of  the  physical  properties  that  produce  the
broadening of the  iron line in an X-ray  pulsar like {\saxj}
is invaluable for accretion theories of fast rotators,
as it  gives constraints on one  of the key parameters,  the radius at
which  the disc  terminates as  the magnetosphere  starts to  lift off
matter towards  the magnetic poles.   Motivated by this, we
 obtained a  63 ks  {\xmm} Target  of  Opportunity (ToO)
observation  of this  source, during  the latest  outburst  in September
2008.

\section{Observation and data reduction}

{\saxj} was found in outburst on 2008 September 21 by {\it RXTE} , and
since then it has been object of an intensive observational campaign.
A preliminary analysis of the 2--10 keV {\it Swift} XRT publicly 
available light curve shows that the outburst had
its maximum around September 24.

{\xmm} observed {\saxj} as a ToO observation for 63 ks on 2008 October 1
(start time MJD 54739.99517), roughly one week after the assumed
outburst peak. The EPIC-pn camera operated in timing mode, to prevent
photon pile-up and to allow an analysis of the coherent and aperiodic timing
behaviour of the source. 
The same observing mode was used for
EPIC/MOS2 CCDs, while EPIC/MOS1 was operated in small window mode, 
and the RGS in
spectroscopy mode. A bright  external flare  in the
background is  present during  the first 2  ks of the  observation, so
that we excluded this time interval from our analysis.

The high source flux saturated  the {\xmm} telemetry rate, so that
the {\xmm} Science Operations Centre  decided to switch off one of
the MOS  cameras (MOS2) in order  to allocate more band  to the EPIC--pn
instrument, roughly 35 ks after  the beginning of the observation. The
EPIC--pn  spectrum showed no  noticeable difference  before and  after the
MOS-2  turning  off, and  the  whole  EPIC--pn  data set  is  therefore
considered for our analysis. The  MOS spectra were severely affected by
pile up, and we did not consider them for our analysis.

Data were extracted and reduced using SAS v.8.0.0. We produced a calibrated
EPIC--pn event list through the \textit{epproc} pipeline, using the most
updated calibration tools (\textit{epfast} tool). We extracted the source
spectrum selecting a 13 pixel wide stripe around the source position (RAWX=37)
equivalent to $53".3$ (which should encircle more than $90\%$ of the energy up
to 9 keV \footnote{see {\xmm} Users handbook, issue 2.6, available at
http://xmm.esac.esa.int/external/xmm\_user\_support/.}), and considered only
PATTERN$\leq$4 and FLAG=0 events. The background spectrum was extracted in the
RAWX=4--15 and RAWY=2--198 CCD region.

The EPIC--pn energy channels were grouped with a compression factor of three so
as not to oversample the instrument energy resolution. The count rate observed
in the EPIC--pn CCD showed a slight increase during the observation from $\sim$
700 c/s, at the beginning of the observation, to 780 c/s at the end of the
observation.

RGS  data were  processed using the \textit{rsgproc} pipeline  to produce
calibrated event  lists, first and  second order spectra  and response
matrices.  The net count rates observed  by the RGS1 and RGS2 in their
first order, which are the only considered here, are 17.8 c/s and 20.0
c/s respectively. RGS spectra were binned in order to have at least 25
counts/bin.

To have a benchmark of  the pn response, we also analysed, using the
standard  pipelines, a  1.1 ks  Swift XRT  observation taken  in window
timing mode (ObsId 00030034033), which started on MJD 54740.60118, and
is therefore simultaneous with the XMM-Newton pointing.
The X-ray spectral package we use to model the observed emission is HEASARC 
{\it XSPEC} v.12.4.

\section{Spectral analysis and results}

\begin{table}
\begin{minipage}[t]{\columnwidth}
\caption{Fitting parameters of the 0.6-11.0 keV combined RGS + EPIC--pn spectrum (centre) and of the 1.4--11.0 keV EPIC--pn spectrum (right) of {\saxj}. The continuum was modelled  by a \texttt{diskbb} (DBB), a \texttt{bbodyrad} (BB) and a power law (PL). Errors on each parameter are quoted at the 90\% confidence level, as for the rest of the paper.}
\label{tab}
\centering
\renewcommand{\footnoterule}{}  
\begin{tabular}{lrr}
\hline \hline
Parameter & 0.6--11.0 keV & 1.4--11.0 keV \\
\hline
nH ($10^{22}$ cm$^{-2}$)\footnote{The absorption column in the EPIC-pn spectrum was varied in the range indicated by the RGS--EPIC-pn spectrum.} & $0.214^{+0.002}_{-0.003}$ & $0.214^{+0.002}_{-0.003}$ \\
kT$_{in}$ (keV) & $0.184^{+0.004}_{-0.002}$& $0.24\pm 0.01$ \\
N$_{DBB}$ ($\times 10^3$) & $36.7^{+8.2}_{-8.0}$&$15.1^{+2.7}_{-2.2}$ \\
kT$_{BB}$ (keV) & $0.36\pm0.01$&$0.40\pm0.02$ \\
R$_{BB}$ ($d_{3.5}$ km) & $9.4^{+0.4}_{-0.3}$ &$6.8^{+1.0}_{-0.9}$ \\
$\alpha$ & $2.079^{+0.006}_{-0.008}$ &$2.078^{+0.007}_{-0.008}$  \\
\hline
Edge E (keV) & $0.871$ (frozen) & \\
$\tau$ O VIII ($\times 10^{-2}$) & $4.6^{+0.5}_{-0.6}$ & \\
\hline
LineE (keV) & $6.47^{+0.07}_{-0.08}$&$6.43^{+0.07}_{-0.09}$ \\
Emissivity index &$-2.2_{-0.3}^{+0.2}$ &$-2.3^{+0.3}_{-0.2}$ \\
R$_{in}$ ($GM/c^2$) & $6.0^{+6.0}$ &$8.7^{+3.7}_{-2.7}$ \\ 
R$_{out}$ ($GM/c^2$) & $210_{-70}^{+150}$ &$207^{+111}_{-80}$ \\
i $(^{\circ})$ & $>58$ &$> 67$ \\
$N_{K_{\alpha}}$ ($\times 10^{-4}$) & $7.2\pm0.9$ &$7.2^{+0.9}_{-0.7}$ \\
EW (eV) & $123^{+29}_{-31}$ &$121^{+20}_{-16}$ \\
\hline
F$_{2-10}$ (erg/cm$^{2}$/s)\footnote{Absorbed flux in the 2.0--10.0 keV band}& $7.58(1)\times 10^{-10}$ & $7.57(3)\times 10^{-10}$  \\
F$_{bol}$ (erg/cm$^{2}$/s)\footnote{Unabsorbed flux extrapolated to the 0.05--150 keV band}  & $4.9\times10^{-9}$ &$4.5\times10^{-9}$ \\
\hline
$\chi^2_{red}$ & $4293.8/3154$ &$809.9/639$ \\
\hline
\end{tabular}
\end{minipage}
\end{table}

In our spectral analysis, we have first considered the combined spectrum of the RGS
1-2 (0.6--2.0 keV) and of the EPIC-pn (2.0--11.0 keV) data. The choice of the lower
bound of the RGS is due to the presence of several calibration residuals at
$\sim 0.5$ keV we chose not to model. The model we have first focused on is composed
of a multi temperature disc emission (\texttt{diskbb}) and a power law. We model
the effect of the interstellar photoelectric absorption using the \texttt{phabs}
component. The addition of another (single temperature) soft component is
strongly required by the data, as the chi squared improves by
$\Delta\chi^2=1104$ with the addition of just two parameters for 3156 degrees of
freedom (d.o.f.). An edge at 0.871 keV, identified with O VIII absorption, is
present in the RGS data at a high significance ($\Delta\chi^2=160$ for the
addition of 1 parameter). A line at $6.5$ keV also clearly emerges in the
residuals (see Fig.{\ref{fig:iron}), at the energy of K$\alpha$ fluorescent
emission of neutral or mildly ionized iron. We obtain a $\Delta\chi^2=399$ for
the addition of 3 parameters when we include a Gaussian to fit this component.
The line is indeed very broad, as we obtain $\sigma=1.1\pm0.2$ keV. This
broadness, together with the apparent asymmetric profile, leads us to consider a
model in which the iron emission is broadened by relativistic motion near 
the NS. We therefore find that a \texttt{diskline} model
\citep{F89} fits  this feature well, giving an inner disc radius consistent with
the range expected from accretion theories onto a pulsar ($6-15$ R$_g$ in the
case of {\saxj}, see discussion). This model, together with residuals, is
plotted in Fig.{\ref{fig:spettro}} and the obtained best fit parameters are
listed in the middle column of Table {\ref{tab}}. 

The final chi squared of the fit in the 0.6--11.0 keV is still quite large and
this is due to the presence of several features, especially in the RGS data,
which are probably of calibration origin (e.g. at 0.7 and 0.85 keV, see
residuals of Fig.{\ref{fig:spettro}}), and which we choose not to fit. 
Considering that the main goal of the observation (and of this Letter) 
is the study of the iron $K\alpha$ line, in order to 
check the stability of the
obtained model, and to prevent spurious effects from a possible wrong
cross-calibration between the instruments, we have also analysed the EPIC-pn dataset
alone, in a bandwidth extending to the lowest energy allowed. We are forced to
limit the energy range to the 1.4--11.0 keV band, as below large residuals
($>6\sigma$) appear with respect to any reasonable continuum model, 
as we could check with a simultaneous fit with the {\it Swift} XRT dataset. 
We fit two evident absorption features at 1.8 and 2.2 keV, 
which are probably of instrumental origin as they lie near the Si K and 
Au M edges,  which are known to influence the effective 
area determination if incorrectly calibrated.
However, we take advantage of the low energy spectral information
given by the RGS data, by taking the absorption column to vary in the range 
determined by the best fit in the whole RGS/pn energy range.  
A quick glance at Table {\ref{tab}} confirms how the best fit parameters
do not change significantly fitting the pn alone. This is especially true
for the hard power-law component and the iron emission line parameters, 
which are indeed perfectly consistent with those obtained from the fit in the
0.6--11 keV band. 

We note that the \texttt{diskline} model gives a slightly better fit 
for the Fe--K$\alpha$ line than a symmetric Gaussian profile, with an
improvement of $\Delta\chi^2=13$ for the addition of three parameters 
(over 588 d.o.f.) with respect to the Gaussian modelling. 
Together with the evident difficulties in 
explaining the observed broadening of this feature with alternative models 
(see discussion),
this definitely leads us to interpret the line profile as being produced in the
inner regions of the accretion disc. To be conservative we discuss our
results in the light of both the best fits presented in Table 1, which 
we remark differ significantly only for the soft components, which themselves are 
less constrained by the EPIC--pn alone.

\begin{figure}
\resizebox{\hsize}{!}{\includegraphics[angle=-90]{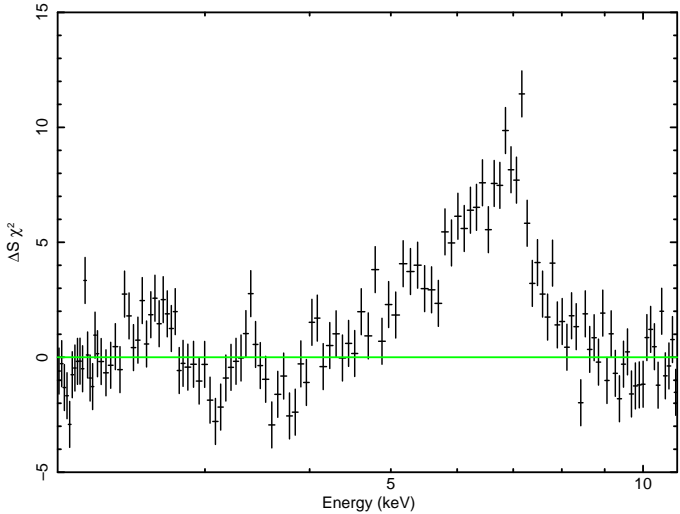}}
\caption{Residuals, in units of $\sigma$, of the 1.4--11.0 keV 
EPIC-pn spectrum, with respect to the best fit continuum model 
(right column of Table \ref{tab}) when the iron line is not included in the fit.}
\label{fig:iron}
\end{figure}

\begin{figure}
\resizebox{\hsize}{!}{\includegraphics[angle=-90]{1401fig2.eps}}
\caption{$0.6-11.0$ keV {\xmm} combined RGS+EPIC-pn spectrum of {\saxj} (top) 
and residuals, in units of $\sigma$, with respect to the
best fit model given in the central column of Table {\ref{tab}}. 
Blue refers to the EPIC-pn, red to RGS1 and green to RGS2 data and residuals. The model components are also plotted: the power law as a magenta solid line, \texttt{diskbb} as a light blue dashed line, \texttt{bbodyrad} as an orange dash-dotted line and the K$\alpha$ \texttt{diskline} component as a dotted magenta line.}

\label{fig:spettro}
\end{figure}

To obtain an estimate of the unabsorbed bolometric flux we extrapolate the  
model of Table {\ref{tab}} to the 0.05--150 keV band. We are confident 
about our estimate, as values of $\alpha\simeq 2$ are easily observed 
in {\saxj} (Gilfanov et al. 1998). 
We finally remark that the 2--10 keV flux 
estimate we obtain is in good accord with the one that can be derived 
by fitting our model to the {\it Swift} XRT spectrum 
($F_{XRT}(2-10)=(7.8 \pm 0.2)\times 10^{-10}$ erg cm$^{-2}$ s$^{-1}$).

Once the orbital motion is corrected \citep{PTR08}, the well known
coherent pulsation is easily detected in the pn data at a frequency 
$\nu=400.9752102$ Hz, consistent with the 2005 value (H08). 
The power spectrum at frequencies lower than $300$ Hz is presented in 
Fig.{\ref{fig:ps}}, and is in line with the typical power spectra 
observed for this source \citep[see e.g.][]{vanstraaten_03}.
Three Lorentzians, with central frequencies of 0.92(5) Hz, 1.56(18) Hz, and 34.7(6) Hz,
respectively, are used to model the timing features we observe. The
first two (labelled A and B in the figure) describe the broad bump already
observed below 10 Hz during other outbursts of this source.
The third Lorentzian, labelled C, could be tentatively identified with the
component observed during the 2002 outburst by \cite{vanstraaten_05}, who named
it L$_h$.

\begin{figure}
\includegraphics[width=8.4cm]{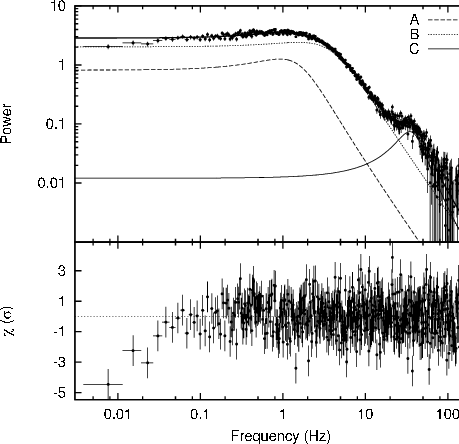}
\caption{(Top panel) White-noise subtracted PSD of the 63 ks {\xmm} observation
         of {\saxj}. The time series was sampled at 125 $\mu$s corresponding to
         a Nyquist frequency of 4000 Hz. The PSD was obtained averaging about
         500 power spectra of $2\times 10^{20}$ temporal bins ($\simeq$ 133 s)
         each. The white-noise (1.97423(25)) was estimated by fitting the
         spectrum between 2 and 3 kHz with a constant. The spectrum was
         logarithmically rebinned using a factor 1.01. Spurious frequencies
         related to the 167 Hz instrumental signal \citep[see][]{Juett_03} were
         removed from the spectrum. The three components labelled as A, B and C
         are the Lorentzians used to fit the spectrum. (Bottom panel) Residuals
         in units of $\sigma$.}
\label{fig:ps}
\end{figure}

\section{Discussion}

The spectrum of the AMSP {\saxj} observed by {\xmm} is of extremely good
quality. For the first time this opens  the possibility of exploiting its 
high sensitivity and spectral resolution capabilities  to study the 
innermost regions of the accretion disc around a rapidly rotating pulsar. 
The estimate of the inner disc radius we obtain from the broadening of 
the iron K$\alpha$ fluorescent line is invaluable 
to test the theories of accretion onto a fast object. 
This can be used to constrain the truncation disc radius, 
which is expected to lie within the centrifugal boundary represented 
by the corotation radius, as indicated by
the presence of pulsations at the known $401$ Hz frequency in the {\xmm} light
curve. Moreover it will be useful for an
estimate of the lever arm of the torque exerted on the NS because of
accretion, when a timing solution covering the whole outburst will be available.

Despite that the measured asymmetry of the iron line does not exclude a Gaussian
profile on a statistical basis, we propose an interpretation of its large 
broadening in terms of the relativistic motion of the reflecting material 
in the space-time bent by the gravitational influence of the NS. 
Disc reflection in fact seems the only viable option to explain the 
large broadening and the low ionization state measured for the emitting
material, in a source whose broadband spectrum is dominated by a hard (high
temperature, low optical depth) Comptonized component (GDB).

For X-ray pulsars, the inner  disc radius, $R_d$, has to meet some observational
constraints implied by the presence of pulsations. $R_{d}$ must
be larger than the NS surface, $R_*\simeq 10$ km depending on the equation of
state, and not much larger than the corotation radius, defined as the point
where the magnetosphere rotation equals that of an assumed Keplerian disc,
$R_c=(GMP^2/4\pi^2)^{1/3}=31 m_{1.4}^{1/3}$ km for {\saxj}, where m$_{1.4}$ is the NS mass in 
units of $1.4$ M$_{\odot}$. 
If not met, the latter condition implies that the magnetic field is not able 
to capture matter because it spins more rapidly than the disc, therefore 
resulting in a centrifugal barrier. Theoretical understanding and 3D MHD 
simulations (\citet{R08} and references therein) showed how the corotation 
condition is not to be taken strictly, as the magnetic field is able to 
release enough angular momentum to the disc to make it leap over the 
centrifugal barrier. In any case an inner disc radius well beyond the 
corotation radius would imply an efficient propeller
ejection of matter, which seems not to be the case for {\saxj} (which is 
observed to show pulsations during this observation, and therefore to accrete 
matter onto the magnetic poles).

The estimate we made for the inner disc radius from the iron line broadening is
$R_{in}=18.0^{+7.6}_{-5.6}$ m$_{1.4}$ km. While the lower bound essentially overlaps  
the NS radius,  the upper limit  we give at a 90\% confidence level 
as $R_{in}<25.6$ m$_{1.4}$ km is in agreement with the expectation that 
the inner disc radius is inside the corotation boundary.
Also in the case of a more massive 2.0 M$_{\odot}$ NS, our upper limit is still
of the order of R$_c$, $R_{in}^{(2.0)}/R_c^{(2.0)}<1.05$. This measurement thus
well fits in the small zone available for a fast rotating NS to efficiently 
accrete matter.

The line modelling holds if we consider both the broader 0.6--11.0 keV
RGS+EPIC--pn spectrum and the restricted 1.4--11 keV bandwidth. The spectral
information in the former is nevertheless fundamental to constrain the 
absorption column to {\saxj}. 
The value we obtain is significantly larger than the average Galactic
value ($0.13 \times 10^{22}$ cm$^{-2}$). This could in principle indicate the
presence of additional neutral absorbers in the proximity of the source, possibly
revealing mass lost from the system during the Roche Lobe overflow \citep{dS08}. The
presence of an evident O VIII edge in the RGS data also indicates the 
presence of ionized absorbing matter along the line of sight.

The measured continuum is in good agreement with the expectations for these 
kinds of sources (GDB, \cite{GP05}). We clearly detect two soft components, 
interpreted as coming from the accretion disc and from the NS surface
(or a good fraction of it), as well as
a dominant hard component, which is well described by a simple power-law in 
the XMM-Newton energy band. 
To derive the bolometric X-ray luminosity, we use our estimate of the
0.05--150.0 keV bolometric flux (see Table {\ref{tab}}), and obtain 
$L_X=6.6 \times 10^{36}$ d$_{3.5}^2$, where d$_{3.5}$ is the distance 
to the source in units of $3.5$ kpc \citep{G06}. 
This estimate is in line with the bolometric flux measured for previous 
outbursts of this source (see e.g. \citet{H08}). In
the hypothesis the X-ray luminosity well tracks the mass accretion rate
onto the NS,
we obtain a mass accretion rate of $\mdot\simeq 5.6 \times 10^{-10}$ d$_{3.5}^2$
R$_6$ m$_{1.4}^{-1}$ M$_{\odot}$/yr at this stage of the outburst, an estimate however affected by the
uncertainties implied by its derivation from a tighter bandwidth spectrum.

The parameters measured for the soft components (see Table {\ref{tab}}) can be
used to obtain further constraints on the geometry of the emitting regions. The
normalization of the multi temperature accretion disc black body (\texttt{diskbb}), 
 $N_{DBB}$, can be used to obtain an estimate of the apparent
inner radius $r_{in}=0.35\sqrt{N_{DBB}/\cos{i}}$ d$_{3.5}$ km, which is indeed
of the order of $R_{in}$ (see e.g. \citet{K98}). In our case we obtain different
values from the fit in the two considered spectral bands, $r_{in}\simeq$40--70 $
({\cos{i}})^{-1/2}$ km, and large uncertainties are still present in the value 
of the inclination angle $i$ of the system ($i \ga 60^\circ$ as derived from 
the diskline fit). Despite still being affected by large uncertainties,
the range of values we derive for the inner disc radius from
the \texttt{diskbb} component is still quite reasonable, and in agreement
with the measure derived from the line fitting.

The single temperature blackbody component we detect with a high statistical
significance in our spectral fits most probably comes from the NS
surface (or a good fraction of it) as indicated by the radius of the 
blackbody emitting region which is $6-10$ km (for a distance to the source 
of 3.5 kpc, see Table 1), in line with the expectations of a hotspot origin.

Thus, in this Letter we have reported the results obtained from 
very high quality {\xmm} spectra. Although some systematic features are 
still present in these spectra (due to the high statistics), the best fit 
model is stable and the best fit parameters do not dramatically depend on
the energy band (or instrument) in which the spectral fit is performed.
The most important result we report here concerns the observation of a 
broad emission line at $\sim 6.5$ keV. The most probable interpretation of 
this feature is reflection of the hard Comptonization spectrum in the 
innermost accretion disc. The inner disc radius derived in this way
confirms the expectation that the accretion disc is truncated 
inside the corotation radius. This is the first time such an estimate 
can be made for a fast rotating X-ray pulsar, and confirms the basic 
 elements of theories of accretion onto these objects.

During the preparation of this manuscript we became aware that 
a $\sim 40$ ks {\it Suzaku} observation also detected the iron line object of this
paper \citep{CK08}. Although the parameters they found for the continuum are
slightly different to those reported in this Letter, the conclusions 
stated there about the line modelling confirm our findings.

\begin{acknowledgements}
We thank N.Schartel, who made possible this ToO observation in the Director Discretionary Time, M.Diaz Trigo M.Guainazzi, and all the {\xmm} team who performed and supported this observation. We thank the referee for the prompt reply, and the Editorial board of A$\&$A for the perfect assistance in the pubblication of this paper. A.P. would also like to thank E.Bozzo and E.Piconcelli for support and useful discussions. 
\end{acknowledgements}

\bibliographystyle{aa}
\bibliography{1401}
\end{document}